\documentclass{epl}
\usepackage{a4,graphicx,rotating,subfigure,amsmath,amsfonts,amssymb,delarray} 
\renewcommand{\vec}[1]{\boldsymbol #1}
\newcommand{\e}{\text{e}}
\newcommand{\im}{\text{i}}
\title{Temperature driven crossover phenomena in the correlation lengths of the one-dimensional t-J model}
\shorttitle{Temperature driven crossover phenomena etc.}
\author{J. Sirker \and A. Kl\"umper}
\shortauthor{J. Sirker \etal}
\institute{Theoretische Physik I, Universit\"at Dortmund - Otto-Hahn-Str.~4, D-44221 Dortmund, Germany}

\begin{document}
\bibliographystyle{phys}
\maketitle

\begin{abstract}
We describe a modified transfer matrix 
renormalization group (TMRG) algorithm and apply it to calculate thermodynamic properties of the one-dimensional t-J model. At the supersymmetric point we compare with Bethe ansatz results and make direct connection to conformal field theory (CFT). In particular we study the crossover from the non-universal high T lattice into the quantum critical regime by calculating various correlation lengths and static correlation functions. Finally, the existence of a spin-gap phase is confirmed. 
\end{abstract}

The t-J model is one of the most fundamental systems of strongly correlated electrons. The two-dimensional version has attracted much attention because it is believed that it describes the basic interactions in the copper-oxygen planes of high-$T_c$ superconductors. 
For the one-dimensional (1D) t-J model much progress has been achieved using various analytical and numerical techniques \cite{OgataLuchini,HellbergMele,BaresBlatter,KawakamiYang,JuettnerKluemper,ChenLee,KobayashiOhe,NakamuraNomura}. 
At the supersymmetric point (SUSYP) $J/t=2$ the model is solvable by the Bethe ansatz (BA)
and ground state properties as well as the excitation spectra have been
obtained exactly \cite{BaresBlatter}. Because the two critical excitations of
spin and charge type are separated, they can be described by two independent
$c=1$ Virasoro algebras. By a combination of finite-size results from the BA and CFT it is therefore also possible to calculate the critical exponents of algebraically decaying correlation functions (CF's) confirming
Tomonaga-Luttinger liquid (TLL) properties \cite{KawakamiYang}. 
By numerical calculations it was shown that the phase diagram consists of a TLL phase for small $J/t$ and a phase separated state for $J/t$ large \cite{OgataLuchini,HellbergMele}. In between a spin-gap (Luther-Emery) phase was conjectured for low densities \cite{OgataLuchini} and confirmed a few years later by different methods \cite{ChenLee,KobayashiOhe,NakamuraNomura}. However, the results for the spin-gap phase strongly depend on the applied methods and the phase boundaries are controversial. Much less is known about thermodynamics of the 1D t-J model. Only at the SUSYP thermodynamic quantities have been obtained by the BA \cite{JuettnerKluemper,OkijiSuga}.

\begin{figure}
\onefigure[scale=0.5]{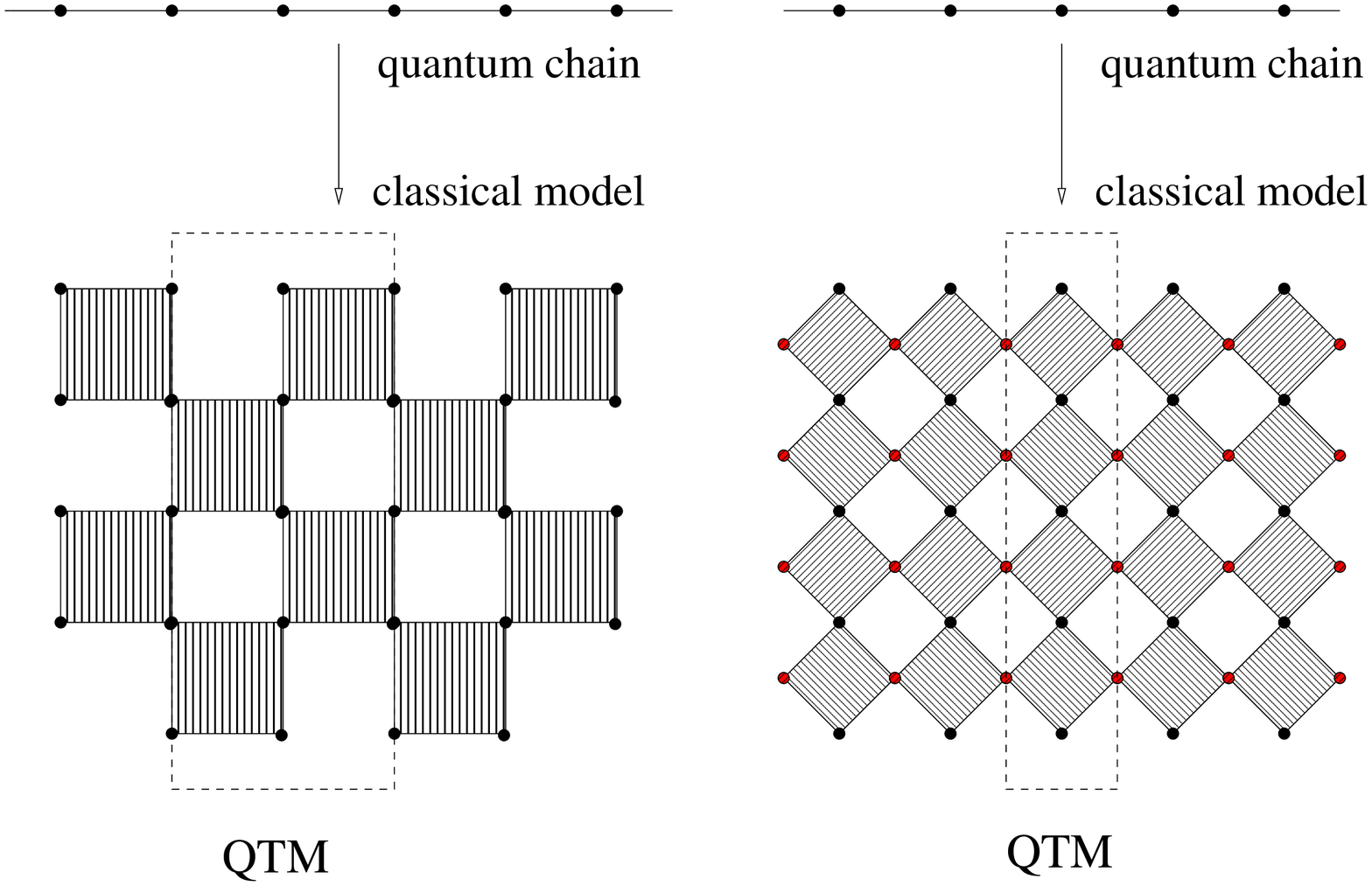}
\caption{The left part shows the usual Trotter mapping of the 1D quantum chain to a 2D classical model with checkerboard structure where the vertical direction corresponds to imaginary time. All lattice points of the classical model belong to the physical lattice at different imaginary time steps and the QTM is a two-column transfer matrix. The right part shows the alternative mapping as described in the text. The classical model has alternating rows and additional lattice points in a mathematical auxiliary space. The QTM in this formulation is only one column wide. In both figures the shaded plaquettes denote the same Boltzmann weight.}
\label{f.1}
\end{figure}

To study thermodynamic properties away from the SUSYP the TMRG provides a powerful numerical tool. This method is particularly suited, because the thermodynamic limit is performed exactly and it has been applied successfully to various 1D systems before \cite{TMRG}. The original idea was to decompose a Hamiltonian $H$ with nearest neighbour interactions into even ($H_e$) and odd parts ($H_o$). By applying the Trotter formula the partition function is expressed as
\begin{equation}
\label{e.0}
Z = \text{Tr}\;\e^{-\beta H} = \lim_{M \rightarrow \infty}\text{Tr}\left\{\left[\e^{-\epsilon H_e}\e^{-\epsilon H_o}\right]^M\right\} 
\end{equation}
with $\epsilon =\beta/M$, $\beta$ being the inverse temperature and $M$ an integer Trotter number leading to a classical model where the column-to-column transfer matrix (QTM) has checkerboard structure (see fig.~\ref{f.1}).
However, this transfer matrix is unnecessarily wide as the repeat length of the classical system is $2$. The consequence are several
disadvantages: (1) The wavevector $k$ of a CF is not
uniquely determined
, i.e.~this transfer matrix cannot distinguish between $k$ and $k+\pi$; (2) the calculation of CF's is complicated, because even and odd as well as distance $1$ have to be treated separately; (3) the cost of computer memory is unnecessarily large. 

We therefore applied a different Trotter-Suzuki mapping leading to a 
classical lattice with alternating rows (see fig.~\ref{f.1}) where the partition function is given by
\begin{equation}
\label{e.1}
Z = \lim_{M \rightarrow
  \infty}\text{Tr}\left\{\left[T_1(\epsilon)T_2(\epsilon)\right]^{M/2}\right\} \quad
  \mbox{with} \quad T_{1,2}(\epsilon) = T_{R,L}\exp\left[-\epsilon H + \mathcal{O}(\epsilon^2)\right]
\end{equation}
and $T_{R,L}$ being the right- and left-shift operators, respectively. Such a mapping is often used in the context of exactly solvable models and is in general applicable for 1D systems with nearest neighbour interactions. 
In this formalism the QTM is formulated for a single column, thus overcoming the disadvantages of the checkerboard decomposition. In both transfer matrix approaches the thermodynamic limit for a fixed Trotter number $M$ is performed exactly, because the free energy of the chain with infinite length is given solely by the largest eigenvalue $\Lambda_0$ of the QTM, $T_M$. Within the alternative mapping the free energy is given explicitly by
\begin{equation}
\label{e.2}
f_{\infty,M} = -T \ln \Lambda_0 \; ,
\end{equation}
where $\Lambda_0$ is unique and a real, positive number for all temperatures. Since a vanishing gap between leading and next leading eigenvalue indicates a phase transition, such a degeneracy is not possible for a 1D quantum system at finite temperature. The QTM is a real but non-symmetric matrix with a block structure due to conserved quantities (here spin and charge or equivalently the number of up-spin electrons $N_\uparrow$ and down-spin electrons $N_\downarrow$). The two-point CF is given by
\begin{equation}
\label{e.3}
\left< A_0 A_r \right> = \sum_\alpha M_{\alpha} \e^{-r/\xi_\alpha} \e^{\im k_\alpha r} \quad , \quad \xi_\alpha^{-1} = \ln \left| \frac{\Lambda_0}{\Lambda_\alpha} \right| \quad , \quad k_\alpha = \arg \left( \frac{\Lambda_\alpha}{\Lambda_0} \right) 
\end{equation}
with $M_\alpha$ being matrixelements that can be directly evaluated and correlation lengths (CL's) $\xi_\alpha$ and wavevectors $k_\alpha$ which are given by nextleading eigenvalues $\Lambda_\alpha$ of the QTM. Note that several $\xi$ with the same $k$ appear in the asymptotic expansion. In the structure factor each term yields a (measurable) Lorentz function with center at $k_\alpha$, height $\sim M_\alpha \xi_\alpha/\pi$ and width $\sim 2/\xi_\alpha$. The sharpest peak corresponds to the leading instability towards the onset of long range order and hence, a crossover in the leading CL indicates a change of the nature of the long range order. 
In addition, it is also possible to calculate imaginary time correlations
$G(r,\tau)$ directly within the TMRG algorithm \cite{Peschel}. The fatal point
is that the analytical continuation 
is an ill-posed problem leading to unreliable
results. What is calculated without fundamental problems are static CF's defined by
$G(r,z=0) = \int_0^\beta d\tau \; G(r,\tau)$.
This CF can be expressed by the largest eigenvalue and the corresponding left and right eigenvectors $|\Psi_0^L\rangle,\,|\Psi_0^R\rangle$ of the QTM:
\begin{equation}
\label{e.5}
G(r,z=0) = \frac{\epsilon}{M \Lambda_0^{r+1}} \langle \Psi^L_0 | \tilde{T}_M T_M^{r-1} \tilde{T}_M | \Psi^R_0 \rangle \quad\mbox{with}\quad \tilde{T}_M = \sum_{k=0}^M T_M(A_{\epsilon \cdot k})
\end{equation}
for distances $r\geq 1$, where $T_M(A_{\epsilon \cdot k})$ denotes the usual transfer matrix $T_M$ with the considered operator $A$ added at imaginary time position $\tau=\epsilon \cdot k$. The static autocorrelation $G(r=0,z=0)$ has to be treated separately in a similar way.

The infinite DMRG algorithm is used to increase the Trotter number $M$ being equivalent to a decrease of the temperature $T$. The numerical algorithm is quite similar to the one described in Ref.~\cite{Peschel} for the checkerboard QTM and will therefore not be explained here in detail.
In the following, we use a grandcanonical description of the t-J model, where the Hamiltonian is given by
\begin{equation}
\label{e.7} 
H \!=\! \sum_i \left\{ -t\sum_\sigma {\it P} (c^{\dagger}_{i,\sigma}
  c_{i+1,\sigma} + c^{\dagger}_{i+1,\sigma} c_{i,\sigma}) {\it P} + J \left( \vec{S}_i \vec{S}_{i+1} - \frac{n_i n_{i+1}}{4}\right) - h \vec{S}^z_i - \mu\, n_i \right\}
\end{equation}
with ${\it P}$ being the projection operator onto the Hilbert-subspace without double occupancy.
\begin{figure}
\twoimages[scale=0.29]{fig2.eps}{fig3.eps}
\caption{Deviation of the free energies calculated by means of TMRG in comparison to the exact results for different temperatures. The lines are guides to the eye. The inset shows the BA results (symbols) plotted on the TMRG results (lines). The three chemical potentials correspond to high ($\mu=-0.7$), medium ($\mu=-1.4$) and low ($\mu=-1.9$) electron density at low temperatures.}
\label{f.2}
\caption{The main figure shows the spin susceptibilities, the inset the charge susceptibilities (compressibilities), where again lines denote the numerical results and symbols the exact ones. The symbols at $T=0$ denote in both graphs the CFT results (see text below).}
\label{f.3}
\end{figure}
Note, that the density is a function of chemical potential and temperature $n= n(\mu,T)$. To check the accuracy of both TMRG algorithms, we have calculated several thermodynamic quantities at the SUSYP and compared with BA results.  
With $100$ states retained in the DMRG and $\epsilon=0.05$, the
accuracy of the free energy in comparison to BA is of the order
$10^{-4}$ if $T/t>0.1$ and of the order $10^{-3}$ for temperatures down to
$T/t = 0.01$ (see fig.~\ref{f.2}) while the deviations between the two different algorithms
are always an order of magnitude smaller. 
By applying a small magnetic field or a small change in the chemical potential, we have calculated spin ($\chi_s$) and charge susceptibilities ($\chi_c$) at the SUSYP as shown in fig.~\ref{f.3}. Again the numerical results are in good agreement with the exact results. Expressing the susceptibilities as a sum of two-point CF's, the high temperature asymptotics is obtained as $\chi_s\sim 1/6T$ and $\chi_c\sim 2/9T$, also consistent with numerics. CFT yields the zero temperature values $\chi_s = 1/(2\pi v_s)$ and $\chi_c=\xi_c(Q)^2/(\pi v_c)$ with $v_{s,c}$ being the velocities of the spinon and holon excitations, respectively, and $\xi_c(Q)$ being the dressed charge. We have calculated the velocities as well as the dressed charge by BA leading to the ground-state susceptibilities shown in fig.~\ref{f.3}. Note, that $v_c$ vanishes for the limiting cases $n\rightarrow 0$ or $n\rightarrow 1$ leading to a divergent compressibility at $T=0$. All analytical results together show that the TMRG algorithm yields accurate results from the high-temperature limit to temperatures of the order $T/t=0.01$ corresponding to $2000$ renormalization steps. Finally, we want to mention that neither the described algorithm nor the one based on the checkerboard decomposition ever breaks down after performing only a few RG-steps as described in Ref.~\cite{RiceTroyer} although we have varied $J/t$ from the free fermion point to the phase separated state and the number of states between $24$ and $300$. 
\begin{figure}
\oneimage[width=1.0\textwidth,height=0.35\textheight]{fig4.eps}
\twoimages[width=0.49\textwidth,height=0.19\textheight]{fig5.eps}{fig6.eps}
\caption{Temperature dependence of the leading d-d and s-s CL's for $J/t=2.0$ and $\mu=-1.4$. Each curve corresponds to one eigenvalue of the QTM and the respective wavevectors may take identical or different values. The triangle up (star) gives the zero temperature result from CFT for the non-oscillating s-s (d-d) and the square that for the $4k_F$ part of the d-d. The triangle down at zero temperature denotes the CFT value for the $2k_F$ part of s-s and d-d, whereas the triangles down at $T/t=0.1$ are given by CFT plus logarithmic corrections as described in the text. The inset shows the wavevectors $k$ in the case of incommensurate oscillations. The circles denote the values for $k=\pm 2 k_F$ at zero temperature as expected from CFT. 
In the temperature range shown in the inset, the density $n(\mu,T)$ varies only between 0.51 and 0.52.}
\label{f.4}
\caption{Static s-s CF at $T=2.0$ (left graph) showing $\pi$-oscillations and at $T=0.1$ (right graph) showing incommensurate oscillations. The dotted lines denote envelopes corresponding to the exponential decay. The straight lines are guides to the eye.}
\label{f.5}
\end{figure}

The supersymmetric t-J model belongs to the TLL universality class with
gapless spin and charge excitations meaning there is a quantum critical point
at $T=0$. The linear dispersion of the critical excitations leads to universal
low-temperature properties like a quadratic temperature dependence of the free
energy $f\sim e_0-aT^2$, a linear regime in the specific heat and CL's diverging as $\xi\sim 1/T$. More insight has been gained from
conformal invariance describing the t-J model by two independent Virasoro
algebras with central charges $c=1$ but different Fermi velocities
$v_{c,s}$. The critical exponents of algebraically decaying correlation
functions have been obtained by the BA due to the relation of finite
size energy gaps and scaling operators in CFT \cite{KawakamiYang}. By the
usual conformal mapping of the complex plane onto a cylinder these results are
easily extended to finite temperatures. However, universality is only given in
the so called {\it quantum critical regime} determined by $T\ll t$. If this
condition is not fulfilled the properties are non-universal and dependent on
the microscopic Hamiltonian. To study the crossover from the non-universal
{\it high T lattice} into the quantum critical regime, we focus on the temperature dependence of CL's and static CF's.  

Respecting the selection rules, the density-density (d-d) and longitudinal
spin-spin (s-s) CL's are in the block of the QTM with
unchanged quantum numbers ($\Delta N_\uparrow = \Delta N_\downarrow = 0$). To
distinguish between them, the matrixelement $M_\alpha$ in eq. (\ref{e.3}) has
to be calculated explicitly. 
The leading d-d and s-s CL's (times temperature) and the corresponding
wavevectors for $J/t=2.0$ and $\mu=-1.4$ ($n_{T\rightarrow 0} \approx 0.52$)
with zero magnetic field are shown in fig.~\ref{f.4}. The non-universal regime
is characterized by several crossovers between the CL's driven by
temperature. In the high temperature limit all CL's show commensurate
oscillations (i.e.~$k=0,\pi$) with a leading $\pi$-oscillating s-s CL and a
leading non-oscillating d-d CL. In the low-temperature regime, $T\ll t$, no
crossovers occur and we expect universal properties. Here the oscillations of a s-s
and a d-d CL are incommensurate with wavevectors depending on
temperature. Note, that the crossovers to incommensurate oscillations are shifted to lower temperatures with increasing particle density and do not occur in the Heisenberg limit $n=1$. 
Because the Fermi momentum $k_{F\uparrow(\downarrow)}$ for up
(down) spin electrons at zero temperature is given by
$k_{F\uparrow(\downarrow)}=\pi (n\pm 2m)/2$ where $m$ is the magnetization, we
identify these CL's as the $2k_F$-oscillating parts. Using the CFT results for
the critical exponents of the algebraically decaying CF's at
$T=0$ \cite{KawakamiYang} and applying the conformal mapping onto the
cylinder, we obtain 
\begin{equation}
\label{e.8}
\xi = \frac{1}{2\pi T \left(\frac{x_c}{v_c}+\frac{x_s}{v_s}\right)} =:
\frac{\gamma}{T}
\end{equation}
with the {\it scaling dimensions $x_{c,s}$}. For the non-oscillating part of the s-s (d-d) CL we receive
$\gamma = v_s/2\pi$ ($\gamma = v_c/2\pi$) whereas $\gamma = 2\, v_c/\{\pi
  (2\frac{v_c}{v_s}+\xi_c(Q)^2)\}$ for both $2k_F$-oscillating parts. In the
  d-d correlation there is also a $4k_F$-oscillating part with $\gamma =
  v_c/\{2 \pi \xi_c^2(Q)\}$, however, this CL is so small that we have not
  calculated it numerically. For the non-oscillating parts we find a good
  agreement with the numerical results, but the $2k_F$ s-s and d-d CL's are
  not equal as expected from CFT. This is a consequence of different
  logarithmic corrections which are not directly accessible within BA but known from TLL theory \cite{GiamarchiSchulz}. The multiplicative
  term $\ln^{-3/2} r$ ($\ln^{1/2} r$) for the $2k_F$ part of the d-d (s-s) correlation at
  $T=0$ can be regarded as an effective, distance dependent correction of the
  scaling dimension $x$ at finite temperature
\begin{equation}
\label{e.8.1}
x' = x - \frac{1}{2}\frac{\ln(\ln^\alpha r)}{\ln r}\qquad\quad (\alpha =-3/2\, , \, 1/2)\, ,
\end{equation}
where the relevant length scale $r$ is the correlation length at the
considered temperature. This correction leads to an excellent agreement
with the numerical results (see triangles down in fig.~\ref{f.4}).
Following only the largest s-s CL a sharp crossover at a critical temperature $T_c\approx 0.8$ from $\pi$-oscillations to non-oscillations occur. However, the matrixelement belonging to the non-oscillating CL is rather small and we expect the $2k_F$ CL dominating the asymptotics of the spin CF for $T\ll T_c$. This crossover is also visible when the static spin CF is calculated explicitly. At $T=2.0$ the CF shows $\pi$-oscillations (see left part of fig.~\ref{f.5}) and a fit $\langle S^z(0)S^z(r)\rangle \sim \exp(-r/\xi)\cos(kr+\delta)$ gives perfect agreement for $\xi$ and $k$ within errors of the order $10^{-4}$ with the direct calculation of CL's. On the other hand, incommensurate oscillations dominate at $T=0.1$ as shown in the right part of fig.~\ref{f.5}. Again the fit values coincide rather precisely with the directly calculated values. Although these crossovers at well defined finite temperatures are visible in physical quantities (i.e.~two-point CF's), they have nothing to do with phase transitions. We want to point out that any thermodynamic quantity derived from the free energy is an analytic function at finite $T$. Phase transitions and corresponding singularities only occur at $T=0$. However, quantities describing the {\it asymptotics of correlation functions} (i.e.~CL) may show non-analyticities even at finite temperature.   

As already mentioned in the introduction a spin-gap or Luther-Emery (LE) phase is expected for values of $J/t$ near the phase separated state at least for small densities \cite{OgataLuchini,HellbergMele,ChenLee,KobayashiOhe,NakamuraNomura}. 
Although the TMRG algorithm is not suited to determine the precise phase boundaries, because a spin gap $\Delta$ is only visible at temperatures $T<\Delta$, 
 we want to show exemplarily and free of any additional assumptions the existence of such a phase.    
\begin{figure}
\twoimages[width=0.49\textwidth,height=0.2\textheight]{fig7.eps}{fig8.eps}
\caption{Spin susceptibilities for two different parameter sets $(J/t,\mu)$, both showing a spin gap of the order $\Delta\sim 0.05$.}
\label{f.6}
\caption{Correlation lengths at $J/t=2.9$ and $\mu=-2.13$. Note, that the oscillations of all shown correlation lengths are commensurate over the entire temperature range.}
\label{f.7}
\end{figure}
The spin susceptibilities for two different values of $J/t$ are shown in
fig.~\ref{f.6}, where the chosen chemical potentials correspond in both cases
to $n\approx 0.2$ in the low-temperature limit. The quadratic dispersion of a
1D gapped system leads to $\chi\sim \exp(-\Delta/T)/\sqrt T$ for the low-temperature susceptibility. Using this function for a fit of the numerical data, we find $\Delta=0.05\pm 0.01$ in both cases. Another proof of LE properties is given by the calculation of s-s and d-d CL's (see fig.~\ref{f.7}). In the low-temperature limit the non-oscillating s-s CL seems to be finite ($T\cdot\xi\rightarrow 0$) whereas the non-oscillating d-d CL is not affected and diverges as $\xi\sim 1/T$. However, we are not able to present numerical data for lower temperatures, which could support this scenario further. 
 
Summarising, we have described a modified TMRG algorithm based on a novel
Trotter-Suzuki mapping where the corresponding QTM is only one column
wide. The advantages are unambiguously determined wavevectors $k$, a
simplified calculation of CF's and a reduction in required computer memory. By calculating 
thermodynamic quantities
at the SUSYP for different chemical
potentials and comparison with BA, we have shown that the numerical
algorithm yields very accurate results down to temperatures of $T/t=0.01$
($2000$ RG steps). In particular, we have investigated temperature driven
crossovers in the correlation lengths and have compared the numerical results
in the quantum critical regime with CFT where logarithmic corrections turned
out to be important. For $J/t\approx 3.0$ and low densities a spin gap was
estimated from the spin susceptibilities proving the existence of a LE phase
what was further supported by the calculation of s-s and d-d CL's.


\acknowledgments
JS thanks Ch. Scheeren for communicating his results from BA and is also grateful for valuable discussions with A. Kemper. This work is supported by the DFG in SP1073.

\end{document}